\documentclass{article}
\usepackage{spconf,amsmath,graphicx}

\usepackage{enumitem}
\setlist{nosep, leftmargin=14pt}

\usepackage{mwe} % to get dummy images
\usepackage{amsfonts}
\usepackage[colorlinks=true,linkcolor=blue, citecolor=blue, urlcolor=blue]{hyperref}

\title{Predicting fluorescent labels in label-free microscopy images with pix2pix and adaptive loss in Light My Cells challenge 
}

\name{Han Liu$^{1}$, Hao Li$^{1}$, Jiacheng Wang$^{1}$, Yubo Fan$^{1}$, Zhoubing Xu$^{2}$, Ipek Oguz$^{1}$}
\address{$^{1}$Vanderbilt University \hspace{25pt} $^{2}$Johnson \& Johnson}

\begin{document}
%\ninept
%
\maketitle

\begin{abstract}
Fluorescence labeling is the standard approach to reveal cellular structures and other subcellular constituents for microscopy images. However, this invasive procedure may perturb or even kill the cells and the procedure itself is highly time-consuming and complex. Recently, in silico labeling has emerged as a promising alternative, aiming to use machine learning models to directly predict the fluorescently labeled images from  label-free microscopy. In this paper, we propose a deep learning-based in silico labeling method for the Light My Cells challenge\footnote{Challenge: https://lightmycells.grand-challenge.org/lightmycells}. Built upon pix2pix, our proposed method can be trained using the partially labeled datasets with an adaptive loss. Moreover, we explore the effectiveness of several training strategies to handle different input modalities, such as training them together or separately. The results show that our method achieves promising performance for in silico labeling. Our code is available at \url{https://github.com/MedICL-VU/LightMyCells}.
\end{abstract}


\section{Introduction}

\begin{figure}[t]
\includegraphics[width=1\columnwidth]{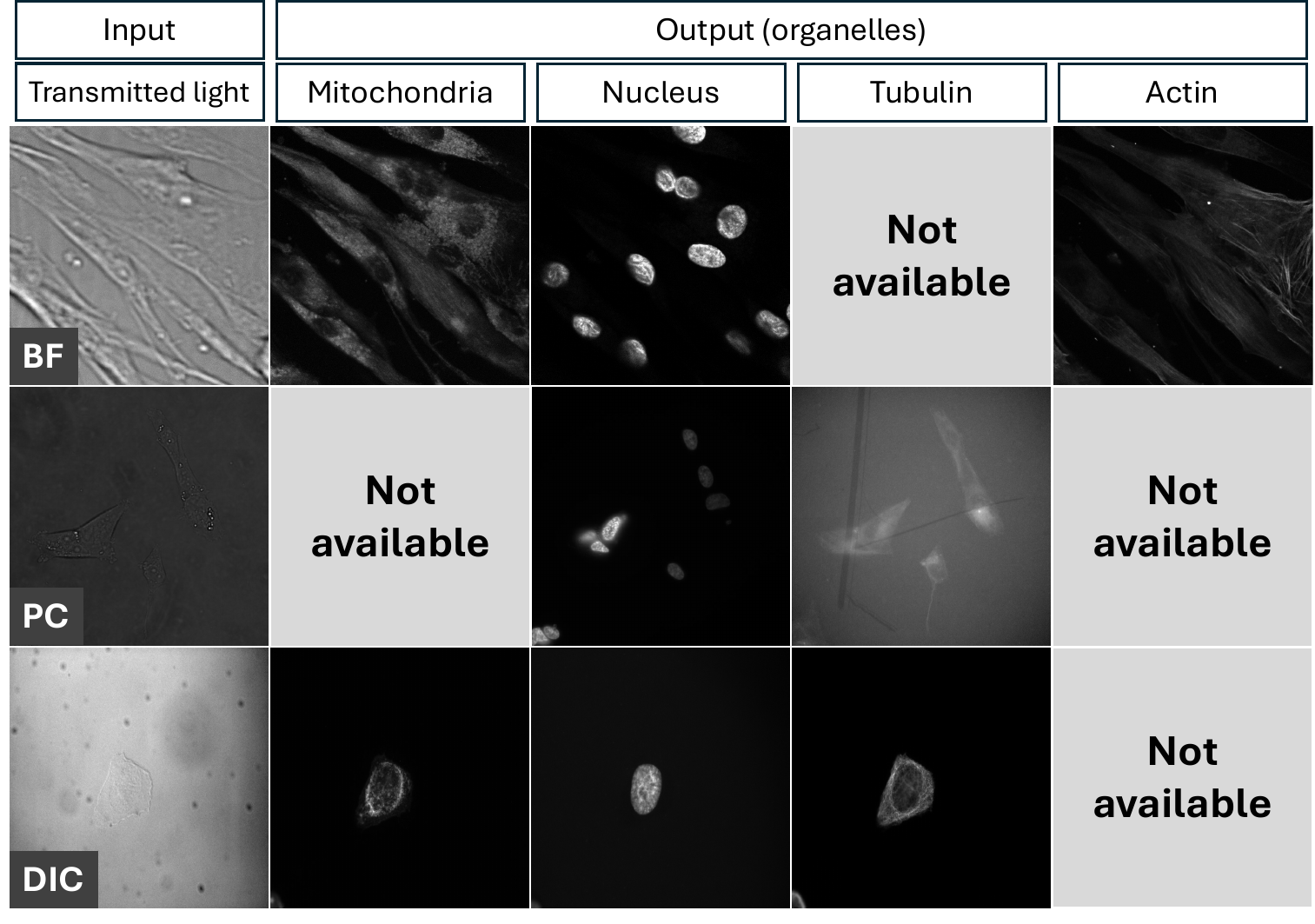}
\centering
\caption{\textbf{Problem formulation of the challenge}. The goal of the challenge is to predict fluorescently labeled images for four organelles (output) from label-free transmitted light microscopy images (input). The input images may have different modalities, (i.e., BF, PC or DIC) and the labels for certain organelles may not be available.} \label{fig1}
\end{figure} 

Recent studies \cite{christiansen2018silico,lee2021deephcs++,cross2022label} show that in silico labeling (ISL), which aims to estimate the fluorescently labeled images in silico directly from the label-free microscopy images, is a promising alternative for traditional fluorescence labeling (FL). In the Light My Cells challenge (Fig. \ref{fig1}), the goal is to predict the best-focused output images of four fluorescently labelled organelles (i.e., mitochondria, nucleus, tubulin and actin) from label-free transmitted light input images, in one of  three modalities, i.e., bright field (BF), phase contrast (PC), and differential interference contrast (DIC).

In this paper, we present our ISL approach based on the well-known image translation method, i.e., pix2pix \cite{isola2017image}. Specifically, we propose to use an adaptive loss to train our model on the partially labeled challenge dataset. Besides, recent studies propose to train unified dynamic networks for heterogeneous datasets, where part of the model parameters are dynamically generated based on some given conditions, such as input modalities \cite{liu2022moddrop++,yao2023unified}, sites \cite{liu2023learning}, body regions \cite{fanct}, etc. In this challenge, we also explore different strategies to handle different input modalities including training separate models, a unified model and a unified dynamic model. Our contributions are summarized as follows.

\renewcommand{\labelitemi}{$\bullet$}
\begin{itemize} 
    \item We develop a pix2pix-based in silico labeling method for heterogeneous microscopy images.
    \item We introduce an adaptive loss to enable training on partially labeled datasets.
    \item We explore different strategies to train with different input modalities, which has not been explored for microscopy images to the best of our knowledge.
\end{itemize}

\begin{figure}[t]
\includegraphics[width=1\columnwidth]{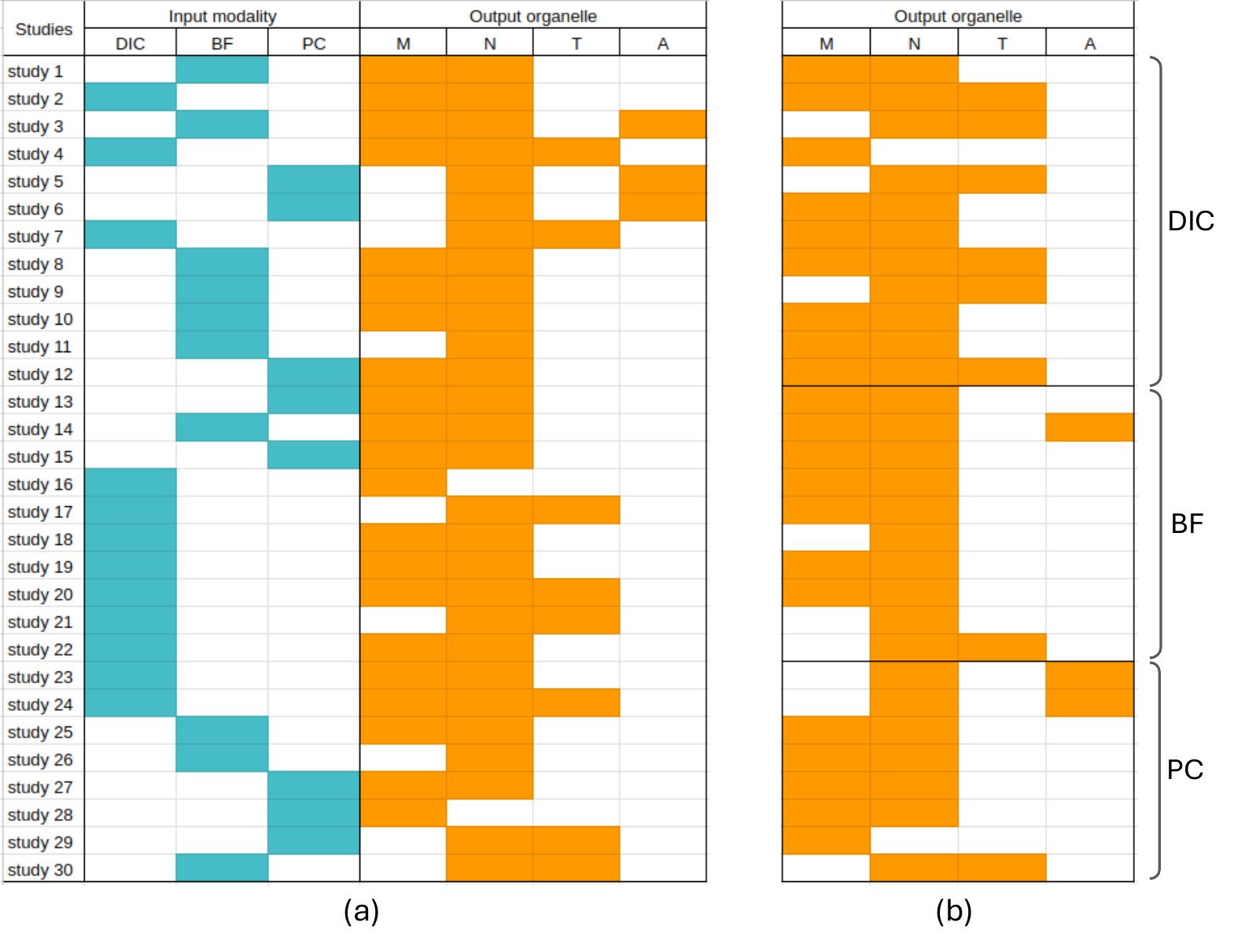}
\centering
\caption{\textbf{Dataset overview}. The challenge dataset consists of 30 sub-datasets collected from different studies. \textbf{(a)}: The available input modalities and labeled organelles for all 30 studies. \textbf{(b)}: The labeled organelles for each input modality (from top to bottom: DIC, BF and PC). M: mitochondria, N: nucleus, T: tubulin, A: actin.} \label{fig2}
\end{figure} 

\section{Materials and Methods}
\subsection{Dataset}
In the Light My Cells challenge, the dataset consists of 30 sub-datasets collected from different studies, including about 57,000 2D microscopy images. The height and width of the 2D images have the range of 512 to 2048. The challenge dataset is highly heterogeneous. First, the microscopy images have a high degree of variability, e.g., magnification, depth of focus, instruments, numerical aperture and modalities (i.e., BF, PC and DIC). Second, the challenge dataset is \textit{partially labeled} \cite{liu2024cosst} because the labels are available only for a few organelles but not for all, as shown in Fig. \ref{fig2}. Third, there is severe class imbalance in this dataset since only a few studies have the labels for tubulin and actin. Even worse, for the DIC modality, none of the studies include the labels of actin.

% As shown in Fig.\ \ref{fig1}, different modalities may have significantly different appearances.

% Since the dataset is partially labeled, the ground truth can be expressed as $y_{i}\in\mathbb{R}^{C_{i}\times H\times W}$ where $C_{i}$ represents the number of labeled organelles for the $i_{th}$ image. 

\subsection{Proposed method overview}
\noindent\textbf{Image preprocessing.} The microscopy images provided by the challenge are stored in an unsigned 16-bit integer format. We rescale all images from the range of $[0, 65535]$ to $[-1, 1]$. 

\noindent\textbf{Class imbalance.} As shown in Fig.\ \ref{fig2}, out of 30 total studies, mitochondria, nucleus, tubulin and actin are included in 21, 28, 8 and 4 studies, respectively. This severe class imbalance may lead to degraded performance, especially for the under-represented classes. To overcome the class imbalance, we propose to sample different organelles \textit{evenly} in each batch. Specifically, we create four lists to record the sample IDs for different organelles. Then we randomly select equal number of samples from each list for each batch.  

\begin{figure}[t]
\includegraphics[width=1\columnwidth]{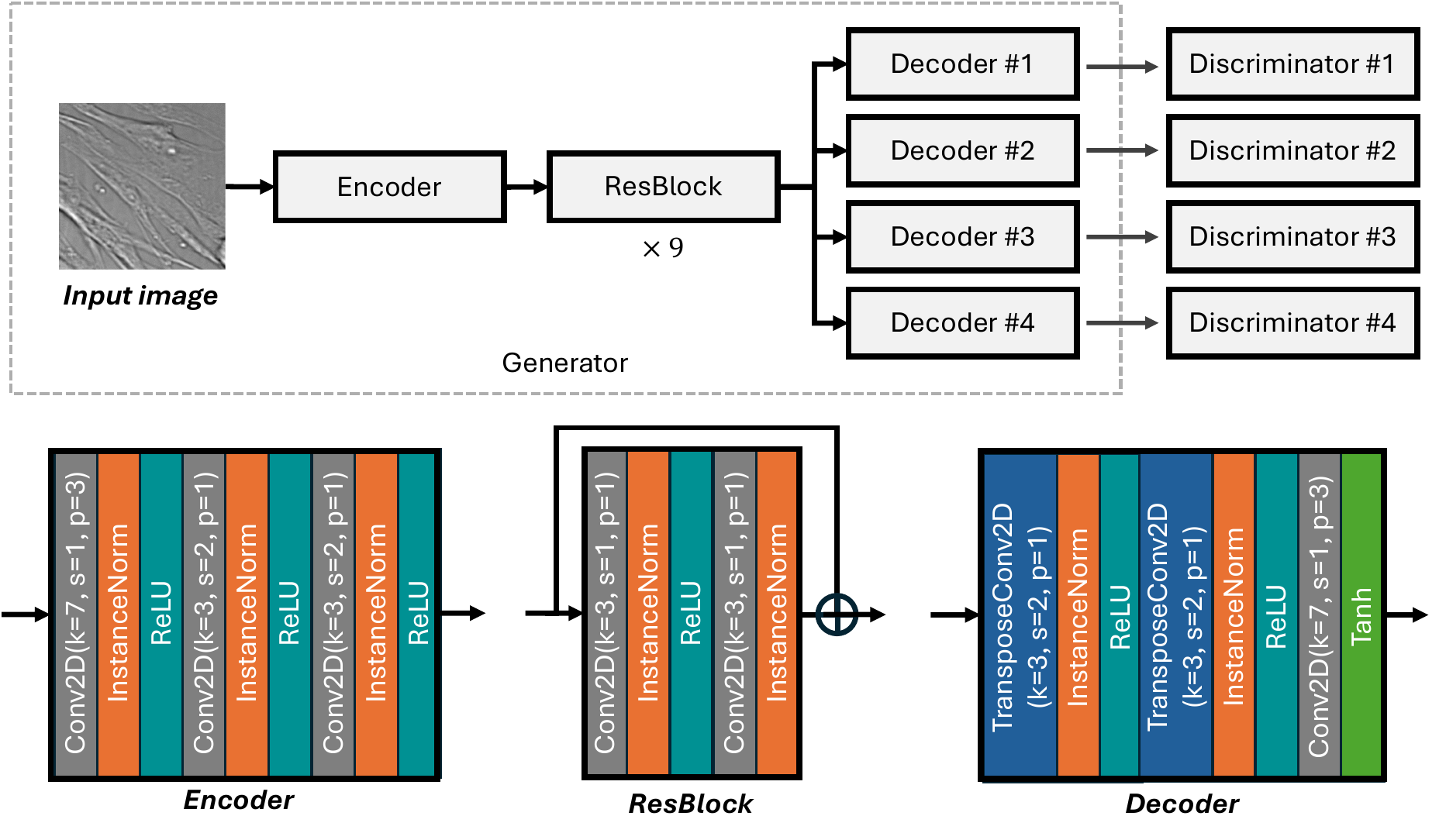}
\centering
\caption{Network architecture of our modified pix2pix model.} \label{fig3}
\end{figure} 

\noindent\textbf{Modified pix2pix model.} Our method is a modified version of pix2pix \cite{isola2017image}, a conditional generative adversarial network (cGAN). Specifically, we propose to use four organelle-specific decoders and discriminators. Compared to a single discriminator for all types of organelles, separate discriminators are assigned with more focused tasks and thus are easier to train. As shown in Fig.\ \ref{fig3}, we use a 2D ResNet (9-block) as the generator and PatchGAN as the discriminator. Since the output images are linearly rescaled to [-1, 1], we apply tanh as the output activation function. During training, the inputs of the model are $512\times512$ patches that are randomly cropped from the original images. For data augmentation, we use random rotation of 90 degrees and random flipping with 50\% probability. During inference, we use a sliding window of the same patch size with an overlap ratio of 0.8. The merged result is finally rescaled back to [0, 65535].

\noindent\textbf{Hyperparameters.} We use the Adam optimizer with an initial learning rate of 0.0002. The learning rate remains the same for the first 150 epochs and linearly decays to 0 for another 150 epochs.
% The batch size is set as 4.

\noindent\textbf{Test-time augmentation.} We apply test-time augmentation (TTA) to improve the final results. Specifically, during inference, random rotation of 90 degrees with 0-3 times are included in the TTA.

\subsection{Adaptive loss for partial label learning}\label{adaptive}
\begin{figure}[t]
\includegraphics[width=1\columnwidth]{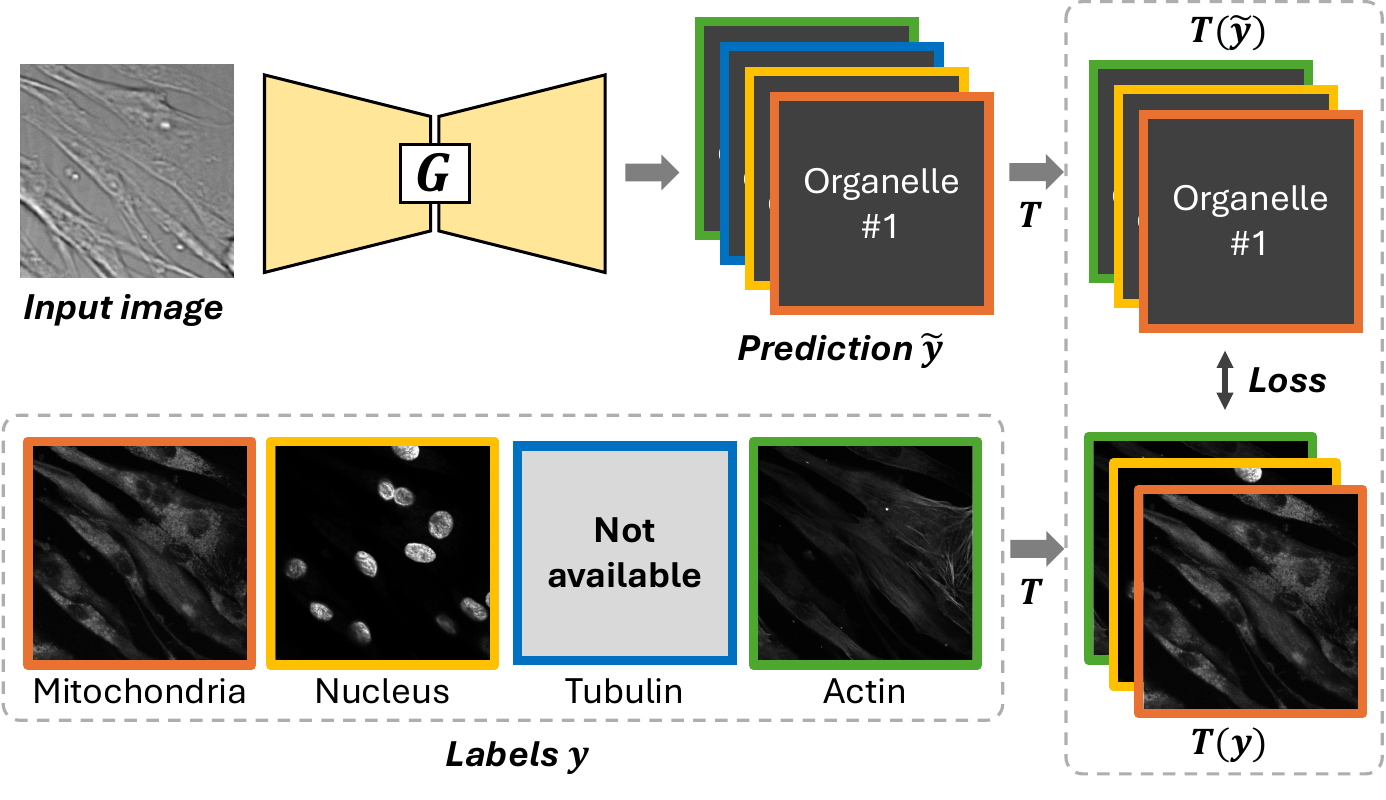}
\centering
\caption{The illustration of the adaptive loss for partial label training. The network prediction $\hat{y}$ and the ground truth $y$ are transformed by removing the unlabeled organelles such that their predictions are excluded from loss computation.} \label{fig4}
\end{figure}

The loss function in vanilla pix2pix assumes the dataset is fully annotated (i.e., all four organelles are labeled), thereby necessitating modifications when the datasets are partially labeled. Inspired by \cite{fang2020multi}, we propose an adaptive loss to enable partial label training, as illustrated in Fig.\ \ref{fig4}. The core idea of the adaptive loss is to only compute the loss for the organelles that have the ground truth. To achieve this goal, we propose to transform the network prediction $\hat{y}$ and the ground truth $y$ by simply removing the predictions of the unlabeled organelles. With the transformed prediction $T(\hat{y})$ and ground truth $T(y)$, the original pix2pix loss $L_{p2p}$ can be applied. The adaptive loss $L_{ada}$ can thus be expressed as: 

\begin{equation}
L_{ada}(\hat{y},y)=L_{p2p}(T(\hat{y}),T(y)) 
\end{equation}

\noindent Specifically, we use a modified pix2pix loss $L_{p2p}$ which consists of a weighted L1 loss and a cGAN loss \cite{isola2017image}:

\begin{equation}
 L_{p2p}=\lambda_{1} L_{L1}\cdot M+\lambda_{2}L_{cGAN}   
\end{equation}

\noindent where $\lambda_{1}$ and $\lambda_{2}$ are set as 100 and 1, respectively. The weighting mask $M$ is created based on the $[2^{th}, 99.8^{th}]$ percentile of the ground truth. We set the weights of the pixels within the percentile range as 1 and the others as 0.1.

% In this challenge, when calculating the evaluation metrics, the predicted results and the ground truth images will be firstly percentile-normalized, i.e., only the pixels within the $[2^{th}, 99.8^{th}]$ percentile are used for evaluation. The reason to use the weighted L1 loss is as follows. Therefore, we aim to make our model focus more on these pixels. 

\subsection{Training strategies}\label{scheme}
\begin{figure}[t]
\includegraphics[width=1\columnwidth]{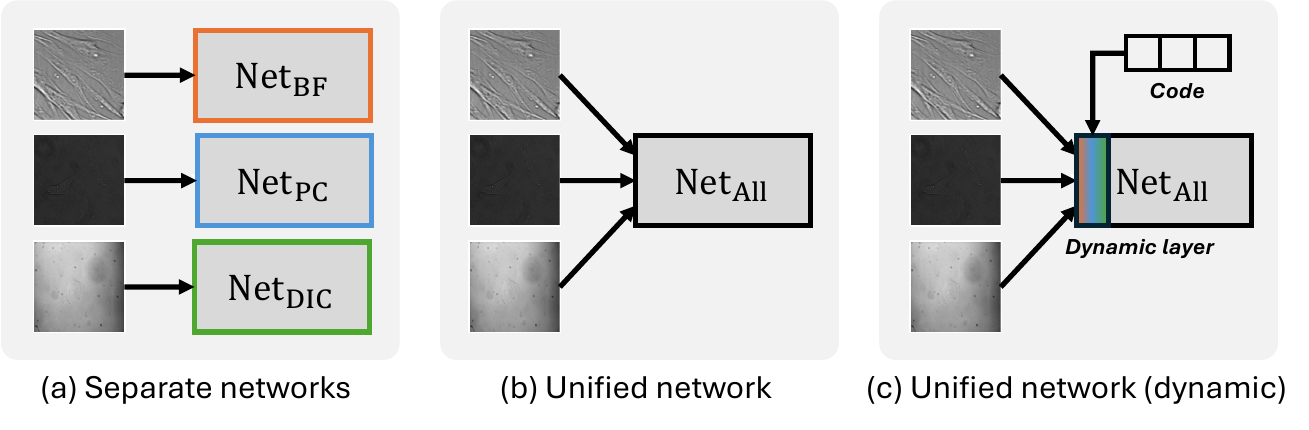}
\centering
\caption{Three training strategies to handle different input modalities.} \label{fig5}
\end{figure}

%\subsection{Separate models vs.\ unified models}\label{scheme}
We observe that the input images with different imaging modalities, i.e., BF, PC and DIC, may have significantly different appearances. To the best of our knowledge, no previous research has explored if the microscopy images with different modalities should be trained together or separately. Specifically, we explore three training strategies to learn from different input modalities, as shown in Fig.\ \ref{fig5}. \textbf{(a) Separate networks.} We build three separate modality-specific networks, where each network is trained using only the images from a single modality. \textbf{(b) A unified network.} We build a single modality-agnostic network, which is trained using all images. \textbf{(c) A unified dynamic network.} We train a dynamic unified network by following \cite{liu2022moddrop++}. Specifically, we replace the first convolutional layer of the network by a dynamic convolutional layer, where the parameters are generated by a one-hot 3-digit modality code. Note that (a) and (c) are valid during inference because the input modalities of testing images can be extracted from metadata. Given the superior performance of UNet++ \cite{zhou2019unet++} in another image translation challenge \cite{zhangtransfer,huijben2024generating}, we use the UNet++ backbone without adversarial training for this experiment.

\begin{figure*}[t]
\includegraphics[width=2\columnwidth]{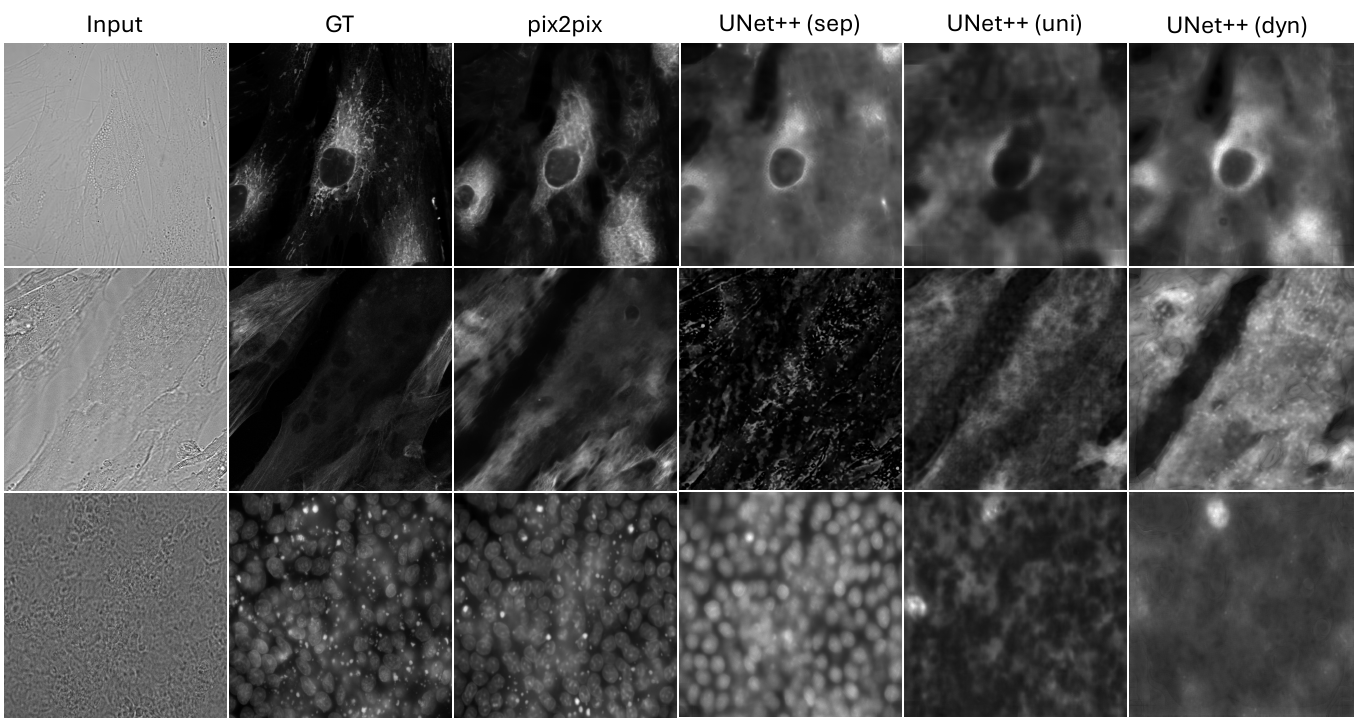}
\centering
\caption{Qualitative results. Each row represents an example of transmitted-light microscopy image (input) being translated into fluorescently labeled images by different methods. Column 4-6 display the impact of different training strategies on UNet++.} \label{fig6}
\end{figure*} 

\section{Experiments and Results}\label{results}
\noindent\textbf{Evaluation metrics.} In this challenge, four metrics are used for quantitative evaluation: (1) Mean Absolute Error (MAE), (2) Structural Similarity Index Measure (SSIM), (3) Pearson Correlation Coefficient (PCC) and (4) Euclidean \& Cosine Distances (E\_dist \& C\_dist). Note that all four metrics are used for mitochondria and nucleus, but only SSIM and PCC are used for tubulin and actin. \textbf{Experiment setup.} For our proposed method, quantitative evaluation is performed by submitting our docker algorithm to the challenge website. Besides, the impact of different training strategies with UNet++ is compared qualitatively.

\subsection{Quantitative results}
We report the quantitative results from the leaderboard of phase 2 in Table \ref{tab1}. Note that our final result is obtained by using three separately trained pix2pix models and an additional unified UNet++ model. The reason is that we find that the pix2pix networks achieve better performance compared to the UNet++ models. However, we cannot obtain reasonable actin prediction by using the model solely trained on the DIC images, simply because no actin are labeled for DIC images, as shown in Fig.\ \ref{fig2}. Therefore, our \textbf{final solution} consists of three separate pix2pix models for general prediction and a unified UNet++ model for actin prediction on DIC images. Compared to our results obtained in phase 1 (only separate pix2pix models are used), the SSIM and PCC of actin prediction are improved from 0.065 to 0.5548 and from 0.003 to 0.5620, respectively. 

% The results will be released by the challenge organizers. 

\begin{table}[h]
\centering
\caption{Quantitative results from the leaderboard of phase 2.}
\label{tab1}
\begin{tabular}{cccccc}
\hline
  & MAE   & SSIM  & PCC   & E\_dist & C\_dist \\
  \hline\hline
M & 0.0958 & 0.6557 & 0.4838 & 229.7230 & 0.3062   \\
N & 0.0798 & 0.7196 & 0.6098 & 195.3898 & 0.2610 \\
T & -   & 0.6699 & 0.5381 & -     & -     \\
A & -   & 0.5548 & 0.5620 & -     & -    \\
\hline
\end{tabular}
\end{table}

\subsection{Qualitative results}\label{qual}
In Fig.\ \ref{fig6}, we display the prediction obtained by different methods on three examples randomly selected from the training set. We observe that \textbf{pix2pix produces the most reasonable prediction compared to the ground truth}. Compared to pix2pix, the results obtained by UNet++ and its variants are less accurate and more blurry, probably due to the lack of adversarial training. Among the different UNet++ training strategies, our results show that \textbf{separate models perform better than the unified  and  dynamic unified models}. Especially in the third row, the unified models produce much worse nucleus prediction compared to the separate models. Our results also align with the findings in \cite{liu2022moddrop++} where the separate networks are shown to be better than unified ones.  

% difference in image translation tasks 

% In our preliminary experiments, we also explore the impacts of training separate and unified models based on UNet++, as shown in column 4-6.

% \textbf{Discussion.} 

\section{Discussion}

% Arguably, there are trade-offs between the separate networks and the unified networks. Separate networks may benefit from learning more focused tasks since the inputs for each network are relatively consistent, but may suffer from training with only a subset of the entire dataset. Unified models may benefit from training on larger datasets but may suffer from learning a more complex task, i.e., learning modality-agnostic representations. 

Arguably, there are trade-offs between the separate networks and the unified networks. For separate networks, they may benefit from learning simpler tasks since the inputs for each network are relatively consistent. However, they may suffer from two drawbacks. First, they can only be trained using a subset of the entire dataset. Second, these modality-specific networks may never learn some input-output mappings due to the partial label issue. For unified models, they may benefit from training on larger datasets but may suffer from learning a more complex task, i.e., encoding different input modalities into modality-agnostic representations. Lastly, for the dynamic unified network, it can be considered as a mixture of the separate and unified networks and may benefit from balancing their trade-offs. Our results show that the separate models outperform the unified models, indicating that the unified models can be more difficult to train than the separate ones when the input modalities have high variability. 

Future directions are in two-fold. First, it is important to explore different normalization techniques for microscopy images. With more standardized images, the pixel-prediction task can be better defined to alleviate the batch-to-batch variability. Second, as proposed by \cite{zhangtransfer}, it is interesting to explore the feasibility of transfer learning for our task, i.e., pretraining with all input modalities (i.e., BF, PC, and DIC) followed by fine-tuning on individual modalities.

\section{Conclusion}
In summary, we presented an in silico labeling method to convert label-free transmitted light microscopy images to fluorescently labeled organelles. Based on pix2pix, our model can be trained on partially labeled challenge dataset with an adaptive loss. The developed method achieves promising results, and our findings show that when the input modalities have high variability, it is more effective to train separate modality-specific models than a single unified model.

\section{Acknowledgements}
This work was supported in part by the National Science Foundation grant 2220401 and the National Institutes of Health grant T32EB021937.

% This work was also supported by the Advanced Computing Center for Research and Education (ACCRE) of Vanderbilt University.

% ------------------------------------------------------------------------- 
\bibliographystyle{IEEEbib}
\bibliography{strings,refs}

\begin{thebibliography}{10}

\bibitem{christiansen2018silico}
Eric~M Christiansen, Samuel~J Yang, D~Michael Ando, Ashkan Javaherian, Gaia Skibinski, Scott Lipnick, Elliot Mount, Alison O’neil, Kevan Shah, Alicia~K Lee, et~al.,
\newblock ``In silico labeling: predicting fluorescent labels in unlabeled images,''
\newblock {\em Cell}, vol. 173, no. 3, pp. 792--803, 2018.

\bibitem{lee2021deephcs++}
Gyuhyun Lee, Jeong-Woo Oh, Nam-Gu Her, and Won-Ki Jeong,
\newblock ``Deephcs++: Bright-field to fluorescence microscopy image conversion using multi-task learning with adversarial losses for label-free high-content screening,''
\newblock {\em Medical image analysis}, vol. 70, pp. 101995, 2021.

\bibitem{cross2022label}
Jan~Oscar Cross-Zamirski, Elizabeth Mouchet, Guy Williams, Carola-Bibiane Sch{\"o}nlieb, Riku Turkki, and Yinhai Wang,
\newblock ``Label-free prediction of cell painting from brightfield images,''
\newblock {\em Scientific reports}, vol. 12, no. 1, pp. 10001, 2022.

\bibitem{isola2017image}
Phillip Isola, Jun-Yan Zhu, Tinghui Zhou, and Alexei~A Efros,
\newblock ``Image-to-image translation with conditional adversarial networks,''
\newblock in {\em Proceedings of the IEEE conference on computer vision and pattern recognition}, 2017, pp. 1125--1134.

\bibitem{liu2022moddrop++}
Han Liu, Yubo Fan, Hao Li, Jiacheng Wang, Dewei Hu, Can Cui, Ho~Hin Lee, Huahong Zhang, and Ipek Oguz,
\newblock ``Moddrop++: A dynamic filter network with intra-subject co-training for multiple sclerosis lesion segmentation with missing modalities,''
\newblock in {\em International Conference on Medical Image Computing and Computer-Assisted Intervention}. Springer, 2022, pp. 444--453.

\bibitem{yao2023unified}
Tianyuan Yao, Nancy Newlin, Praitayini Kanakaraj, Vishwesh Nath, Leon~Y Cai, Karthik Ramadass, Kurt Schilling, Bennett~A Landman, and Yuankai Huo,
\newblock ``A unified learning model for estimating fiber orientation distribution functions on heterogeneous multi-shell diffusion-weighted mri,''
\newblock in {\em International Workshop on Computational Diffusion MRI}. Springer, 2023, pp. 13--22.

\bibitem{liu2023learning}
Han Liu, Yubo Fan, Zhoubing Xu, Benoit~M Dawant, and Ipek Oguz,
\newblock ``Learning site-specific styles for multi-institutional unsupervised cross-modality domain adaptation,''
\newblock {\em arXiv preprint arXiv:2311.12437}, 2023.

\bibitem{fanct}
Yubo Fan, Han Liu, Ipek Oguz, and Benoit~M Dawant,
\newblock ``Ct synthesis with modality-, anatomy-, and site-specific inference,''
\newblock .

\bibitem{liu2024cosst}
Han Liu, Zhoubing Xu, Riqiang Gao, Hao Li, Jianing Wang, Guillaume Chabin, Ipek Oguz, and Sasa Grbic,
\newblock ``Cosst: Multi-organ segmentation with partially labeled datasets using comprehensive supervisions and self-training,''
\newblock {\em IEEE Transactions on Medical Imaging}, 2024.

\bibitem{fang2020multi}
Xi~Fang and Pingkun Yan,
\newblock ``Multi-organ segmentation over partially labeled datasets with multi-scale feature abstraction,''
\newblock {\em IEEE Transactions on Medical Imaging}, vol. 39, no. 11, pp. 3619--3629, 2020.

\bibitem{zhou2019unet++}
Zongwei Zhou, Md~Mahfuzur~Rahman Siddiquee, Nima Tajbakhsh, and Jianming Liang,
\newblock ``Unet++: Redesigning skip connections to exploit multiscale features in image segmentation,''
\newblock {\em IEEE transactions on medical imaging}, vol. 39, no. 6, pp. 1856--1867, 2019.

\bibitem{zhangtransfer}
Yiwen Zhang, Chuanpu Li, Zeli Chen, and Kaiyi Zheng,
\newblock ``Transfer learning and 2.5 d unet++ for cbct-ct synthesis in synthrad2023,''
\newblock .

\bibitem{huijben2024generating}
Evi Huijben, Maarten~L Terpstra, Arthur Galapon~Jr, Suraj Pai, Adrian Thummerer, Peter Koopmans, Manya Afonso, Maureen van Eijnatten, Oliver Gurney-Champion, Zeli Chen, et~al.,
\newblock ``Generating synthetic computed tomography for radiotherapy: Synthrad2023 challenge report,''
\newblock {\em arXiv preprint arXiv:2403.08447}, 2024.

\end{thebibliography}

\end{document}